\documentclass[11pt]{article}
\usepackage{amsmath,amsthm,amssymb}
\date{}
\newtheorem{theorem}{Theorem}
\newtheorem{lemma}{Lemma}

\newtheorem{construction}{Construction}
\newtheorem{conjecture}{Conjecture}
\newtheorem{proposition}{Proposition}

\newtheorem{open question}{Open question}

\title{Determining the Weight Spectrum of the Reed--Muller Codes $RM(m-6,m)$}
\author{ Yueying~Lou\footnote{School of Computer and Electronic Information/School of Artificial Intelligence,
Nanjing Normal University, Nanjing, China. Email: 232212001@njnu.edu.cn.
} \ and \ Qichun~Wang \footnote{School of Computer and Electronic Information/School of Artificial Intelligence,
Nanjing Normal University, Nanjing, China.
 Email: qcwang@fudan.edu.cn.} \footnote{Corresponding author.}}
\begin{document}
\maketitle

\begin{abstract}
The weight spectra of the Reed-Muller codes $RM(r,m)$ were unknown for $r=3,...,m-5$. In IEEE Trans. Inform. Theory 2024, Carlet determined the weight spectrum of $RM(m-5,m)$ for $m\ge10$ using the Maiorana-McFarland construction, where the result was tried to be extended to $RM(m-6,m)$, but many problems occurred and much work needed to be done. In this paper, we propose a novel way of constructing Reed--Muller codewords and determine the weight spectrum of $RM(m-6,m)$ for $m\ge12$, which gives a positive answer to an open question on the weight spectrum of $RM(m-c,m)$ for $c=6$. Moreover, we put forward a conjecture and verify it for some cases. If the conjecture is true, then that open question can be completely solved.
\end{abstract}



\section{Introduction}
An $m$-variable Boolean function is a map from $\mathbb{F}_{2}^{m}$ into the
finite field $\mathbb{F}_{2}$. The $r$--th order Reed--Muller code of length $2^m$ is denoted by $RM(r,m)$. Its codewords are the truth tables (output values) of the set of all $m$-variable Boolean functions of degree $\leq r$. 

The low Hamming weights are well--known for all Reed--Muller codes, and the only Hamming weights in $RM(r,m)$ belonging to the range $[2^{m-r},2^{m-r+1}]$ are of the form $2^{m-r+1}-2^{m-r+1-i}$, where $i\le \max(\min(m-r,r),\frac{m-r+2}{2})$ \cite{Kasami2}. 
This result was extended and all the weights lying between the minimum distance $2^{m-r}$ and $2^{m-r+1}+2^{m-r-1}$ were determined by Kasami, Tokura and Azumi in \cite{Kasami1}.

The weight spectra of the Reed-Muller codes $RM(r,m)$ are also well--known for $r=0,1,2,m-2,m-1,m$ (see e.g. \cite{CP,MacWilliams}). In 2023, Carlet and Sol\'{e} determined the weight spectra of $RM(m-c,m)$ for $c=3,4$ \cite{CP}. Recently, Carlet determined the weight spectrum of $RM(m-5,m)$ for $m\ge10$ using the Maiorana-McFarland construction, where he tried to extend the result to $RM(m-6,m)$, but not succeeded \cite{Carlet1}.

In this paper, we propose a novel way of constructing Reed--Muller codewords and determine the weight spectrum of $RM(m-6,m)$ for $m\ge12$, which gives a positive answer to an open question on the weight spectrum of $RM(m-c,m)$ for $c=6$. Moreover, we put forward a conjecture and verify it for some cases. If the conjecture is true, then that open question can be completely solved.

The paper is organized as follows. In Section~2, the necessary background is established. We determine the weight spectrum of $RM(m-6,m)$ in Section~3, and try to extend the result to $RM(m-c,m)$ in Section~4. We end in Section~5 with conclusion.
\section{Preliminaries}
Let $\mathbb{F}_{2}^{m}$ be the $m$-dimensional vector space over the
finite field $\mathbb{F}_{2}$. We denote by $B_{m}$ the set
of all $m$-variable Boolean functions, from $\mathbb{F}_{2}^{m}$ into $\mathbb{F}_{2}$.

Any Boolean function $f\in B_{m}$ can be uniquely represented
as a multivariate polynomial in
$\mathbb{F}_{2}[x_{1},\cdots,x_{m}]$,
\[
f(x_1,\ldots ,x_m)=\sum_{K\subseteq \{1,2,\ldots ,m\}}a_K\prod_{k\in K}x_k,
\]
 which is called its {\em algebraic normal form}
(ANF). The {\em algebraic degree} of $f$, denoted by $\deg(f)$, is the
number of variables in the highest order term with nonzero
coefficient.

The $r$-th order Reed-Muller code of length $2^m$ is denoted by $RM(r,m)$. Its codewords are the truth tables (output values) of the set of all $m$-variable Boolean functions of degree $\leq r$.

We use $|A|$ to denote the cardinality of the set $A$. Let $f\in B_m$ and
\[
 1_{f}=\{x\in \mathbb{F}_{2}^{m}|f(x)=1\}, \ 0_{f}=\{x\in \mathbb{F}_{2}^{m}|f(x)=0\}.
  \]
 The cardinality $|1_{f}|$ is called the {\em Hamming weight} of $f$, and will be denoted by $wt(f)$. The {\em Hamming distance} between two functions
$f$ and $g$ is the Hamming weight of $f+g$, and will be denoted by $d(f,g)$. It is well--known that $wt(f)$ is odd if and only if $\deg(f)=m$.

Weights of the Reed-Muller code $RM(r,m)$ are the Hamming weights of all its codewords. That is, the Hamming weights of all $f\in B_{m}$ with $\deg(f)\le r$. The weight spectrum of $RM(r,m)$ is the set of all distinct weights.

It is well-known that the weights in $RM(r,m)$ are multiples of $2^{\lfloor\frac{m-1}{r}\rfloor}$ (McEliece divisiblity theorem \cite{Mc}), and its minimum nonzero weight equals $2^{m-r}$. The low Hamming weights are well--known for all Reed--Muller codes which can be seen from the following theorem.
 
\begin{theorem} [\cite{Kasami1}]
Let $w$ be a weight of $RM(r,m)$ in the range $2^{m-r}\le w<2^{m-r+1}$. Let $\alpha=\min(r,m-r)$ and $\beta=\frac{m-r+2}{2}$. Then $w$ is of the form $2^{m-r+1}+2^{m-r+1-i}$, where $1\le i\le \max(\alpha,\beta)$. Conversely, for any such $i$, there is a weight $w$ of that form in the range $2^{m-r}\le w<2^{m-r+1}$.
\end{theorem}
Kasami, Tokura and Azumi determined later in \cite{Kasami2} all the weights lying between
the minimum distance $2^{m-r}$ and $2^{m-r+1}+2^{m-r-1}$.

We use $||$ to denote the concatenation, that is,
\[
(f_1||f_2)(x_1,\ldots,x_m,x_{m+1})=(x_{m+1}+1)f_1(x_1,\ldots,x_m)+x_{m+1}f_2(x_1,\ldots,x_m),
\]
and $ f_1||f_2||f_3||f_4=$
 \[
(x_{m+1}+1)(x_{m+2}+1)f_1+x_{m+1}(x_{m+2}+1)f_2+(x_{m+1}+1)x_{m+2}f_3+x_{m+1}x_{m+2}f_4,
 \]
 where $f_1,f_2,f_3,f_4\in B_m$. There are many results on Boolean functions deduced through using concatenation techniques (see e.g. \cite{Carlet1,Wang,WangT}).
\section{Determining the weight spectrum of the Reed-Muller codes $RM(m-6,m)$}
In this section, we will use the concatenation technique and construct functions of $B_{12}$ with the form $g_0||g_1||g_2||(g_1+g_2+g_3)$ which can achieve all the possible weights of of $RM(6,12)$, where $g_0,g_3\in RM(4,10)$ and $g_1,g_2\in RM(5,10)$.

\begin{lemma} Let $g=g_0||g_1||g_2||(g_1+g_2+g_3)$, where $g_0,g_3\in RM(4,10)$ and $g_1,g_2\in RM(5,10)$. Then $g\in RM(6,12)$.
\end{lemma}
\proof Clearly,
\begin{eqnarray*}
g&=&(x_{11}+1)(x_{12}+1)g_0+x_{11}(x_{12}+1)g_1+(x_{11}+1)x_{12}g_2+x_{11}x_{12}(g_1+g_2+g_3)\\
&=&(x_{11}+1)(x_{12}+1)g_0+x_{11}g_1+x_{12}g_2+x_{11}x_{12}g_3,
\end{eqnarray*}
and the result follows.
\endproof

By the results of \cite{Kasami1,Kasami2}, The set of all weights $<2.5*2^{6}=160$ in $RM(6,12)$ is known, which is shown
as the following lemma.

 \begin{lemma} [\cite{Kasami1,Kasami2}]
 The set of all weights $<160$ in $RM(6,12)$ equals
 \[
 \{0,64,96,112,120,124,126,128,136,144,148,152,154,156,158\}.
 \]
\end{lemma}
Using the Maiorana-McFarland construction, Carlet found that all the even weights from 154 to 190 can be reached except for 166. He then showed that weight 166 can be achieved, but it is a little complex \cite{Carlet1}. We now give a simple way to construct a codeword with the weight 166.

 \begin{lemma} Let $f=x_{i_1}x_{i_2}x_{i_3}x_{i_4}x_{i_5}+x_{i_6}x_{i_7}x_{i_8}x_{i_9}x_{i_{10}}\in B_{10}$. Let $I=\{i_1,i_2,i_3,i_4,i_5\}$ and $J=\{i_6,i_7,i_8,i_9,i_{10}\}$. If $|I\bigcap J|=c$, then $wt(f)=64-2^{c+1}$.
\end{lemma}
\proof Let $f_1=x_{i_1}x_{i_2}x_{i_3}x_{i_4}x_{i_5}$ and $f_2=x_{i_6}x_{i_7}x_{i_8}x_{i_9}x_{i_{10}}$. Clearly,
\begin{eqnarray*}
wt(f)&=&|1_{f_1}\cap 0_{f_2}|+|1_{f_2}\cap 0_{f_1}|\\
&=& (2^{5}-2^{10-|I\cup J|})+(2^{5}-2^{10-|J\cup I|})\\
&=& (32-2^{|I\cap J|})+(32-2^{|I\cap J}|)\\
&=& 64-2^{c+1}.
\end{eqnarray*}
\endproof

By Lemma 3, the set of all weights for functions of the form $x_{i_1}x_{i_2}x_{i_3}x_{i_4}x_{i_5}+x_{i_6}x_{i_7}x_{i_8}x_{i_9}x_{i_{10}}\in B_{10}$ is $ \{0,32,48,56,60,62\}$. Since $62+56+48=166$, it is natural to construct a codeword with weight 166 using the concatenation $0||g_1||g_2||(g_1+g_2)$ satisfying $wt(g_1)=62$, $wt(g_2)=56$ and $wt(g_1+g_2)=48$.

\begin{lemma} Let $g=0||g_1||g_2||(g_1+g_2)$, where $g_1=x_{1}x_{2}x_{3}x_{4}x_{5}+x_{6}x_{7}x_{8}x_{9}x_{10}$ and $g_2=x_{1}x_{2}x_{3}x_{4}x_{5}+x_{1}x_{2}x_{6}x_{7}x_{8}$ Then $g\in RM(6,12)$ and $wt(g)=166$.
\end{lemma}
\proof By Lemma 1, $g\in RM(6,12)$. By Lemma 3, $wt(g_1)=62$, $wt(g_2)=56$ and $wt(g_1+g_2)=48$.
Therefore, $wt(g)=62+56+48=166$, and the result follows.
\endproof

Inspired by the above method to construct the codeword of weight 166, we consider the following construction.

 \begin{construction} Let $g_1=x_{1}x_{2}x_{3}x_{4}x_{5}+x_{i_1}x_{i_2}x_{i_3}x_{i_4}x_{i_5}+x_{i_6}x_{i_7}x_{i_8}x_{i_9}x_{i_{10}}$
 and $g_2=x_{1}x_{2}x_{3}x_{4}x_{5}+x_{i_1}x_{i_2}x_{i_3}x_{i_4}x_{i_5}+x_{i_{11}}x_{i_{12}}x_{i_{13}}x_{i_{14}}x_{i_{15}}$,  where $g_1,g_2\in B_{10}$ and $1\le i_1,\ldots,i_{15}\le 10$. We then construct $g=0||g_1||g_2||(g_1+g_2)$.
\end{construction}

Clearly, $g_1$ and $g_2$ in Construction 1 are with at most three monomials and $g_1+g_2$ is with at most two monomials. By Lemma 1, $g\in RM(6,12)$. Let $J=\{i_6,i_7,i_8,i_9,i_{10}\}$ and $K=\{i_{11},i_{12},i_{13},i_{14},i_{15}\}$. If $|J\bigcap K|=c$, then by Lemma 3, $wt(g_1+g_2)=64-2^{c+1}$. We now determine the weights of functions with three monomials.

\begin{lemma} Let $f=x_{1}x_{2}x_{3}x_{4}x_{5}+x_{i_1}x_{i_2}x_{i_3}x_{i_4}x_{i_5}+x_{i_6}x_{i_7}x_{i_8}x_{i_9}x_{i_{10}}\in B_{10}$. Let $I=\{1,2,3,4,5\}$, $J=\{i_1,i_2,i_3,i_4,i_5\}$ and $K=\{i_6,i_7,i_8,i_9,i_{10}\}$. If $|I\cap J|=c_1$, $|I\cap K|=c_2$, $|J\cap K|=c_3$ and $|I\cap J\cap K|=c_4$, then
\[
wt(f)=2^{c_1+c_2+c_3-c_4-3}-2^{c_1+1}-2^{c_2+1}-2^{c_3+1}+96.
\]
\end{lemma}
\proof Let $f_1=x_{1}x_{2}x_{3}x_{4}x_{5}$, $f_2=x_{i_1}x_{i_2}x_{i_3}x_{i_4}x_{i_5}$ and $f_3=x_{i_6}x_{i_7}x_{i_8}x_{i_9}x_{i_{10}}$. Clearly, $wt(f)=$
\[
|1_{f_1}\cap 0_{f_2}\cap 0_{f_3}|+|1_{f_2}\cap 0_{f_1}\cap 0_{f_3}|+|1_{f_3}\cap 0_{f_1}\cap 0_{f_2}|+|1_{f_1}\cap 1_{f_2}\cap 1_{f_3}|.
\]
By the inclusion-exclusion principle, $|I\cup J\cup K|=15-c_1-c_2-c_3+c_4$. Therefore,
\begin{eqnarray*}
|1_{f_1}\cap 1_{f_2}\cap 1_{f_3}|=2^{c_1+c_2+c_3-c_4-5}.
\end{eqnarray*}
Moreover, we have 
\begin{eqnarray*}
&&|1_{f_1}\cap 0_{f_2}\cap 0_{f_3}|\\
&=&(2^{|J-K-I|}-1)(2^{|K-J-I|}-1)2^{10-|I\cup J\cup K|}+32-2^{5-|J\cap K-I|}\\
&=&(2^{5-c_1-c_3+c_4}-1)(2^{5-c_2-c_3+c_4}-1)2^{c_1+c_2+c_3-c_4-5}+32-2^{5-c_3+c_4}\\
&=&2^{c_1+c_2+c_3-c_4-5}-2^{c_1}-2^{c_2}+32.
\end{eqnarray*}
Similarly,
\begin{eqnarray*}
|1_{f_2}\cap 0_{f_1}\cap 0_{f_3}|&=&2^{c_1+c_2+c_3-c_4-5}-2^{c_1}-2^{c_3}+32\\
|1_{f_3}\cap 0_{f_1}\cap 0_{f_2}|&=&2^{c_1+c_2+c_3-c_4-5}-2^{c_2}-2^{c_3}+32,
\end{eqnarray*}
and the result follows.
\endproof
By Lemma 5, the set of all weights for functions of the form $x_{1}x_{2}x_{3}x_{4}x_{5}+x_{i_1}x_{i_2}x_{i_3}x_{i_4}x_{i_5}+x_{i_6}x_{i_7}x_{i_8}x_{i_9}x_{i_{10}}$ is $ \{32,48,56,60,62,64,68,72,74,76,80\}$. By choosing suitable functions with three monomials, Construction 1 can generate functions whose weights range over all those integers between
154 and 210 that are congruent with 2 modulo 4. 
 \begin{lemma} Take $g_1=x_{1}x_{2}x_{3}x_{4}x_{5}+x_{1}x_{2}x_{3}x_{4}x_{6}+x_{1}x_{2}x_{3}x_{4}x_{7}$ and $g_2=x_{1}x_{2}x_{3}x_{4}x_{5}+x_{1}x_{2}x_{3}x_{4}x_{6}+h$, where $h=x_{5}x_{6}x_{8}x_{9}x_{10}$ or $x_{5}x_{7}x_{8}x_{9}x_{10}$. Then Construction 1 generates the functions with the weights 154 and 158.
\end{lemma}
\proof By Lemma 5, $wt(g_1)=2^5-2^5-2^5-2^5+96=32$ and 
\[
wt(g_2)=\left \{ \begin{array} {l} 2^3-2^5-2^2-2^2+96=64, \ \ \ \ \ \ \ \mbox {if} \ \ h=x_{5}x_{6}x_{8}x_{9}x_{10}, \\
  2^2-2^5-2^2-2^1+96=62, \ \ \ \ \ \ \ \mbox {if} \ \ h=x_{5}x_{7}x_{8}x_{9}x_{10}. \\
\end{array} \right.
\]
By Lemma 3,
\[
wt(g_1+g_2)=\left \{ \begin{array} {l} 64-2^1=62, \ \ \ \ \ \ \ \mbox {if} \ \ h=x_{5}x_{6}x_{8}x_{9}x_{10}, \\
  64-2^2=60, \ \ \ \ \ \ \ \mbox {if} \ \ h=x_{5}x_{7}x_{8}x_{9}x_{10}. \\
\end{array} \right.
\]
Clearly, 32+64+62=158 and 32+62+60=154, and the result follows.
\endproof

 \begin{lemma} Take $g_1=x_{1}x_{2}x_{3}x_{4}x_{5}+x_{1}x_{2}x_{3}x_{6}x_{7}+x_{4}x_{5}x_{8}x_{9}x_{10}$ and $g_2=x_{1}x_{2}x_{3}x_{4}x_{5}+x_{1}x_{2}x_{3}x_{6}x_{7}+h$, where $h$ is a monomial of degree 5. Then Construction 1 can generate the functions with the weights $$\{162,174,178,182,186,190,194,198,202,210\}.$$
\end{lemma}
\proof By Lemma 5, $wt(g_1)=2^2-2^4-2^3-2^1+96=74$. Similarly, the values of $wt(g_2)$ and $wt(g_1+g_2)$ can be computed and their values are given by the following table.
\begin{table}[!h]
 \begin{center}\begin{tabular}{|c|c|c|c|c|c|c|c|c|c|}
  \hline
$h$ & $x_{1}x_{2}x_{3}x_{4}x_{5}$ & $x_{1}x_{4}x_{5}x_{8}x_{9}$ &  $x_{1}x_{2}x_{3}x_{4}x_{8}$  & $x_{1}x_{2}x_{3}x_{4}x_{6}$  &  $x_{1}x_{2}x_{3}x_{8}x_{9}$ \\
  \hline
  $wt(g_2)$ & 32 &  68 & 48 & 48  &  56  \\
  \hline
  $wt(g_1+g_2)$ & 56 &  32 & 56 & 60 &  56  \\
  \hline
  $h$ & $x_{1}x_{2}x_{6}x_{7}x_{8}$ & $x_{1}x_{2}x_{4}x_{5}x_{6}$ &  $x_{1}x_{2}x_{4}x_{6}x_{7}$  & $x_{1}x_{4}x_{6}x_{7}x_{8}$  &  $x_{1}x_{4}x_{5}x_{6}x_{7}$ \\
  \hline
  $wt(g_2)$ & 56 &  64 & 64 & 72 &  80 \\
  \hline
  $wt(g_1+g_2)$ & 60 &  56 & 60 & 56  &  56 \\
  \hline
 \end{tabular}
     \end{center}
     \end{table}        
 Then we can calculate $wt(g_1)+wt(g_2)+wt(g_1+g_2)$, and the result follows.
\endproof

\begin{proposition}
 The set of the weights of 12--variable Boolean functions of the form $0||g_1||g_2||(g_1+g_2)$ contains all those integers between
154 and 214 that are congruent with 2 modulo 4, where $g_1,g_2\in B_{10}$ are homogeneous polynomials of degree 5.
\end{proposition}
\proof By Lemmas 4, 6 and 7, Construction 1 can generate the functions with the weights
\[
\{154,158,162,166,174,178,182,186,190,194,198,202,210\}.
 \]
 Let $g_1=x_{1}x_{2}x_{3}x_{4}x_{5}+x_{1}x_{2}x_{3}x_{6}x_{7}+x_{4}x_{5}x_{8}x_{9}x_{10}$. If we take $g_2=x_{1}x_{2}x_{3}x_{4}x_{5}+x_{1}x_{6}x_{8}x_{9}x_{10}+x_{4}x_{5}x_{8}x_{9}x_{10}$, then $g_1+g_2=x_{1}x_{2}x_{3}x_{6}x_{7}+x_{1}x_{6}x_{8}x_{9}x_{10}$. By Lemmas 3 and 5, $wt(g_1)=74$, $wt(g_2)=76$ and $wt(g_1+g_2)=56$. Therefore, $0||g_1||g_2||(g_1+g_2)$ is of weight 206. Take $g_2=x_{1}x_{2}x_{3}x_{4}x_{5}+x_{1}x_{6}x_{7}x_{8}x_{9}+x_{6}x_{7}x_{8}x_{9}x_{10}$, then $wt(g_2)=62$ and $g_1+g_2=x_{1}x_{2}x_{3}x_{6}x_{7}+x_{1}x_{6}x_{7}x_{8}x_{9}+x_{4}x_{5}x_{8}x_{9}x_{10}+x_{6}x_{7}x_{8}x_{9}x_{10}$. It is easy to be calculated that $wt(g_1+g_2)=78$. Therefore,  $0||g_1||g_2||(g_1+g_2)$ is of weight 214,
and the result follows.
\endproof
Based on Construction 1, we can obtain all weights between 1050 and 1110 that are congruent with 2 modulo 4 by flipping some bits, which can be seen from the following proposition.
\begin{proposition}
Let $g_1=x_{1}x_{2}x_{3}x_{4}x_{5}+x_{1}x_{2}x_{3}x_{6}x_{7}+x_{4}x_{5}x_{8}x_{9}x_{10}$
 and $g_2=x_{1}x_{2}x_{3}x_{4}x_{5}+x_{1}x_{2}x_{3}x_{6}x_{7}+h$,  where $g_1,g_2\in B_{10}$ and $h$ is a monomial of degree 5. Then the set of the weights of 12--variable Boolean functions of the form $0||(g_1+a_1)||(g_2+a_2)||(g_1+g_2+a_3)$ contains all those integers between 1050 and 1110 that are congruent with 2 modulo 4, where $a_1,a_2,a_3\in F_2$.
\end{proposition}
\proof Since $wt(g_1)=74$, we have 
 \[
 wt(0||(g_1+1)||g_2||(g_1+g_2))=876+wt(0||g_1||g_2||(g_1+g_2)).
 \]
  Then by Lemma 7,
 \[
\{1050,1054,1058,1062,1066,1070,1074,1078,1086\}\subseteq \{wt(0||(g_1+1)||g_2||(g_1+g_2))\}.
 \]
 Take $h=x_1x_4x_6x_7x_8$, then $wt(g_2)=72$ and $wt(g_1+g_2)=56$. Hence
 \[
 wt(0||g_1||(g_2+1)||(g_1+g_2))=74+1024-72+56=1082.
 \]
 Take $h=x_1x_2x_3x_4x_8$, then $wt(g_2)=48$ and $wt(g_1+g_2)=56$. Therefore,
\begin{eqnarray*}
 wt(0||g_1||(g_2+1)||(g_1+g_2))&=&74+1024-48+56=1106,\\
 wt(0||g_1||g_2||(g_1+g_2+1))&=&74+48+1024-56=1090.
\end{eqnarray*}
 Take $h=x_1x_2x_6x_7x_8$, then $wt(g_2)=56$ and $wt(g_1+g_2)=60$. Hence,
\begin{eqnarray*}
 wt(0||g_1||(g_2+1)||(g_1+g_2))&=&74+1024-56+60=1102,\\
 wt(0||g_1||g_2||(g_1+g_2+1))&=&74+56+1024-60=1094.
\end{eqnarray*}
Take $h=x_1x_2x_3x_8x_9$, then $wt(g_2)=56$, $wt(g_1+g_2)=56$ and
 \[
 wt(0||g_1||(g_2+1)||(g_1+g_2))=74+1024-56+56=1098.
 \]
 Take $h=x_1x_2x_3x_4x_6$, then $wt(g_2)=48$ and $wt(g_1+g_2)=60$. Therefore,
 \[
 wt(0||g_1||(g_2+1)||(g_1+g_2))=74+1024-48+60=1110,
 \]
and the result follows.
\endproof
Based on Construction 1, we can also obtain all weights between 1056 and 1116 that are congruent with 0 modulo 4.
\begin{proposition}
Let $g_1=x_{1}x_{2}x_{3}x_{4}x_{5}+x_{1}x_{2}x_{3}x_{6}x_{7}+x_{1}x_{7}x_{8}x_{9}x_{10}$
 and $g_2=x_{1}x_{2}x_{3}x_{4}x_{5}+x_{1}x_{2}x_{3}x_{6}x_{7}+h$,  where $g_1,g_2\in B_{10}$ and $h$ is a monomial of degree 5. Then the set of the weights of 12--variable Boolean functions of the form $0||(g_1+a_1)||(g_2+a_2)||(g_1+g_2+a_3)$ contains all those integers between 1056 and 1116 that are congruent with 0 modulo 4, where $a_1,a_2,a_3\in F_2$.
\end{proposition}
\proof By Lemma 5, $wt(g_1)=2^2-2^4-2^2-2^3+96=72$. Take $h=x_1x_2x_3x_4x_7$, then $wt(g_2)=48$ and $wt(g_1+g_2)=56$. Hence
 \begin{eqnarray*}
 wt(0||(g_1+1)||g_2||(g_1+g_2))&=&1024-72+48+56=1056,\\
 wt(0||g_1||(g_2+1)||(g_1+g_2))&=&72+1024-48+56=1104,\\
 wt(0||g_1||g_2||(g_1+g_2+1))&=&72+48+1024-56=1088.
\end{eqnarray*}
 Take $h=x_1x_2x_3x_4x_6$, then $wt(g_2)=48$ and $wt(g_1+g_2)=60$. Therefore,
 \begin{eqnarray*}
 wt(0||(g_1+1)||g_2||(g_1+g_2))&=&1024-72+48+60=1060,\\
 wt(0||g_1||(g_2+1)||(g_1+g_2))&=&72+1024-48+60=1108,\\
 wt(0||g_1||g_2||(g_1+g_2+1))&=&72+48+1024-60=1084.
\end{eqnarray*}
 Take $h=x_1x_2x_4x_5x_8$, then $wt(g_2)=56$ and $wt(g_1+g_2)=56$. Hence,
\begin{eqnarray*}
 wt(0||(g_1+1)||g_2||(g_1+g_2))&=&1024-72+56+56=1064,\\
  wt(0||g_1||(g_2+1)||(g_1+g_2))&=&72+1024-56+56=1096.
\end{eqnarray*}
 Take $h=x_1x_4x_5x_8x_9$, then $wt(g_2)=68$ and $wt(g_1+g_2)=48$. Therefore,
 \begin{eqnarray*}
 wt(0||(g_1+1)||g_2||(g_1+g_2))&=&1024-72+68+48=1068,\\
 wt(0||g_1||(g_2+1)||(g_1+g_2))&=&72+1024-68+48=1076,\\
 wt(0||g_1||g_2||(g_1+g_2+1))&=&72+68+1024-48=1116.
\end{eqnarray*}
Take $h=x_1x_2x_4x_5x_7$, then $wt(g_2)=64$, $wt(g_1+g_2)=56$ and
 \[
 wt(0||(g_1+1)||g_2||(g_1+g_2))=1024-72+64+56=1072.
 \]
 Take $h=x_1x_2x_4x_5x_6$, then $wt(g_2)=64$ and $wt(g_1+g_2)=60$. Hence
 \begin{eqnarray*}
 wt(0||g_1||(g_2+1)||(g_1+g_2))&=&72+1024-64+60=1092,\\
 wt(0||g_1||g_2||(g_1+g_2+1))&=&72+64+1024-60=1100.
\end{eqnarray*}
Take $h=x_1x_4x_5x_6x_8$, then $wt(g_2)=72$ and $wt(g_1+g_2)=56$. Therefore,
 \begin{eqnarray*}
 wt(0||g_1||(g_2+1)||(g_1+g_2))&=&72+1024-72+56=1080,\\
 wt(0||g_1||g_2||(g_1+g_2+1))&=&72+72+1024-56=1112,
\end{eqnarray*}
and the result follows.
\endproof
From the OEIS sequence A146976 \cite{Sloane}, the set of weights in $RM(4, 8)$ contains all $16i$, for $0\le i\le 16$. Moreover, the set of weights in $RM(5,10)$ was determined by Carlet recently which contains all even integers between 72 and 952 \cite{Carlet1}.

\begin{lemma} [\cite{Desaki,Sloane}] The set of weights in $RM(4,8)$ contains all $16i$, where i ranges over the set of consecutive integers from 0 to 16.
\end{lemma}
\begin{lemma} [\cite{Carlet1}] The set of weights in $RM(5,10)$ contains all even integers between 72 and 952.
\end{lemma}
\begin{proposition}
Let $A=\{0,64,96,112,120,124,126,128,136,144,148\}$ and $S$ be the set of all weights in $RM(6,12)$. Then 
 \[
 S=A\cup\{152+2i\}\cup\{2^{12}-a \ | \ a\in A\},
 \]
where i ranges over the set of consecutive integers from 0 to $2^{11}-152$.
\end{proposition}
\proof Consider those functions of the form $g=g_0||g_1||g_2||(g_1+g_2+g_3)$, where $g_0,g_3\in RM(4,10)$ and $g_1,g_2\in RM(5,10)$. By Lemma 1, $g\in RM(6,12)$. Clearly,
\[
wt(g)=wt(g_0)+wt(0||g_1||g_2||(g_1+g_2+g_3)).
\]
By Lemma 8, $wt(g_0)$ can achieve all $64i$, for $0\le i\le 16$. Therefore, by Propositions 1 and 2, $wt(g)$ can achieve all the numbers of the set
\[
\{154+64i\le a\le 214+64i \}\cup\{1050+64i\le a\le 1110+64i\},
 \]
 where $a\equiv2\ (mod \ 4)$ and $0\le i\le 16$. Hence, $S$ contains all those integers between 154 and 2134 that are congruent with 2 modulo 4. If $f\in RM(5,10)$, then $g=0||0||f||f\in RM(6,12)$ and $wt(g)=2wt(f)$. Therefore, by Lemma 9, $S$ contains all those integers between 152 and 1904 that are congruent with 0 modulo 4. Moreover, by Lemma 8 and Proposition 3, $S$ contains all those integers between 1908 and 2048 that are congruent with 0 modulo 4. Therefore, 
 \[
 \{152+2i\ | \ 0\le i\le 2^{11}-76\}\subseteq S.
 \]
 Then by Lemma 2 and $wt(g+1)=2^{12}-wt(g)$ for $g\in RM(6,12)$, 
  \[
 S=A\cup\{152+2i\}\cup\{2^{12}-a \ | \ a\in A\},
 \]
where i ranges over the set of consecutive integers from 0 to $2^{12}-152$. 
\begin{theorem}
Let $A=\{0,64,96,112,120,124,126,128,136,144,148\}$ and $S$ be the set of all weights in $RM(m-6,m)$, where $m\ge 12$. Then
 \[
 S=A\cup\{152+2i\}\cup\{2^{m}-a \ | \ a\in A\},
 \]
where i ranges over the set of consecutive integers from 0 to $2^{m-1}-152$.
\end{theorem}
\proof By Proposition 4, the result is correct for $m=12$. Assuming the result is correct for $m$, we now prove that it is also correct for $m+1$. Let $S$ be the set of all weights in $RM(m-7,m+1)$. Since $0||f\in RM(m-7,m+1)$ for any $f\in RM(m-6,m)$, and $wt(0||f)=wt(f)$, we have
\begin{equation}
 A\cup\{152+2i\ | \ 0\le i\le2^{m-1}-152\}\subseteq S.
 \end{equation}
Let $g_1\in RM(m-6,m)$ with $wt(g_1)=152$. Since $g_1||g_2\in RM(m-7,m+1)$ for any $g_2\in RM(m-6,m)$ and $wt(g_1||g_2)=152+wt(g_2)$, we have
\begin{equation}
\{152+2i\ | \ 2^{m-1}-152 \le i\le2^{m-1}\}\subseteq S.
 \end{equation}
Moreover, $wt(f+1)=2^{m+1}-wt(f)$, for any $f\in RM(m-7,m+1)$. Then by (1) and (2), we have
 \[
 A\cup\{152+2i\}\cup\{2^{m+1}-a \ | \ a\in A\}\subseteq S,
 \]
 where i ranges over the set of consecutive integers from 0 to $2^{m}-152$. Then by \cite{Kasami1,Kasami2} and the fact that the weights in $RM(m-6,m)$ must be even, we have
  \[
 S\subseteq A\cup\{152+2i\}\cup\{2^{m+1}-a \ | \ a\in A\},
 \]
and the result follows.
\endproof
This confirms for $c=6$ the conjecture stated in \cite{CP} about the weight spectrum of $RM(m-c,m)$, which was later presented as an open question since it seemed to be very difficult to make a prediction by further study \cite{Carlet1}. However, by studying Construction 1 and its generalization, we think that the conjecture could be true. So, we prefer to state it as a conjecture.
\begin{conjecture} [Conjecture of \cite{CP}, Open question of \cite{Carlet1}]
Let $c$ be any positive integer. For $m\ge2c$, the weight spectrum of $RM(m-c,m)$ is of the form
\[ 
\{0\}\cup A \cup B \cup C \cup \overline{B} \cup \overline{A} \cup \{2^m\},
\]
where:\\
$A\subseteq [2^c,2^{c+1}]$ is given by Kasami and Tokura \cite{Kasami1},\\
$B\subseteq [2^{c+1},2^{c+1}+2^{c-1}]$ is given by Kasami, Tokura and Azumi \cite{Kasami2},\\
$C\subseteq [2^{c+1}+2^{c-1},2^{m-1}-2^{c+1}-2^{c-1}]$ consists of all consecutive even integers,\\
$\overline{A}$ stands for the complement to $2^m$ of $A$, and $B$ stands
for the complement to $2^m$ of $B$.
\end{conjecture}
\section{Trying to extend the result to $RM(m-c,m)$}
Inspired by the above method to deal with $RM(6,12)$, we consider the following more general construction.
 \begin{construction} Let $g_1=x_{1}x_{2}\cdots x_{m}+x_{i_1}x_{i_2}\cdots x_{i_m}+x_{i_{m+1}}\cdots x_{i_{2m}}$
 and $g_2=x_{1}x_{2}\cdots x_{m}+x_{i_{2m+1}}\cdots x_{i_{3m}}+x_{i_{3m+1}}\cdots x_{i_{4m}}$,  where $g_1,g_2\in B_{2m}$ and $1\le i_1,\ldots,i_{4m}\le 2m$. We then construct $g=0||(g_1+a_1)||g_2||(g_1+g_2+a_2)\in RM(m+1,2m+2)$, where $a_1,a_2\in F_2$.
\end{construction}
By analyzing the weights of those functions in Construction 2 for $m=4,5$, we propose the following more ambitious conjecture, which implies Conjecture 1.
\begin{conjecture}
Let $S$ be the set of all weights generated by Construction 2. Then
\[
\{2^{m+2}+2^m+2i\}\cup\{2^{2m}+2^{m}+2i\}\subseteq S,
\]
where $0\le i<2^{m}$ and $m\ge 4$.
\end{conjecture}
From the discussion in Section 3, Conjecture 1 is true for $m=5$.
It is quite easy to verify the conjecture for $m=4$. In fact, we calculate the weights for those functions of Construction 2 using the computer, and find that there exist many functions achieving  the values of $\{80+2i\}\cup\{272+2i\}$, which can be seen from the following Table 1, where $num(f)$ denotes the number of functions in Construction 2 with the weight $wt(f)$.
\begin{table}[!h]
\caption{Number of $f\in B_{10}$ in Construction 2 with the desired weight}
 \begin{center}\begin{tabular}{|c|c|c|c|c|c|c|c|c|c|}
  \hline
$wt(f)$ & 80 &  82 & 84 & 86  &  88 & 90 &  92 & 94  \\
  \hline
  $num(f)$ & 1426248 &  85248 & 1680384 & 208224  &  2789312 & 351872 &  3152040 & 541824  \\
  \hline
  $wt(f)$ & 96  &  98 & 100 &  102 & 104 & 106  &  108 & 110 \\
  \hline
  $num(f)$ & 3690240  &  516192 & 3553440 &  465024 & 2186472 & 190080  &  940032 & 33696 \\
  \hline
  $wt(f)$ & 272  &  274 & 276 &  278 & 280 & 282  &  284 & 286 \\
  \hline
  $num(f)$ & 2801168  &  323648 & 4203144 &  601632 & 6844464 & 849888  &  7165472 & 916336 \\
  \hline
  $wt(f)$ & 288  &  290 & 292 &  294 & 296 & 298  &  300 & 302 \\
  \hline
  $num(f)$ & 7051536  &  816576 & 5449440 &  629808 & 3956448 & 373984  &  2145576 & 173088 \\
  \hline
 \end{tabular}
     \end{center}
     \end{table}
\begin{proposition}
Conjecture 2 implies Conjecture 1.
\end{proposition}
\proof Let $S$ be the set of all weights of those functions $0||(g_1+a_1)||g_2||(g_1+g_2+a_2)\in RM(m+1,2m+2)$, where $g_1,g_2\in RM(m,2m)$ and $a_1,a_2\in F_2$. If Conjecture 2 is true, then we have
\begin{equation}
\{2^{m+2}+2^m+2i\}\cup\{2^{2m}+2^{m}+2i\}\subseteq S,
\end{equation}
where $0\le i<2^{m}$ and $m\ge 4$. We first prove that the
weight spectrum of $RM(m,2m)$ is the same as that of Conjecture 1. Assuming this is true for $m\le k$, we now consider $m=k+1$. Let $g_0\in RM(k-1,2k)$. Then $wt(g_0)$ can achieve the values $4i$, where $i$ ranges over all the weights of $RM(k-1,2k-2)$. By the assumption, $wt(g_0)$ can achieve $2^{k+1}i$, for $0\le i\le 2^{k-1}$. Hence by (3), $\{2^{k+2}+2^k+2i\}\subseteq S$,
where $0\le i\le 2^{2k}$. Then by the results of \cite{Kasami1,Kasami2} and $wt(g+1)=2^{2k+2}-wt(g)$ for $g\in RM(k+1,2k+2)$, the
weight spectrum of $RM(k+1,2k+2)$ is also the same as that of Conjecture 1. Then similar to the proof of Theorem 2, we can deduce the result.
\endproof
To prove Conjecture 1, by Proposition 5, we only need to determine the weights of some functions in Construction 2 and prove that Conjecture 2 is true. This seems to be feasible if all the desired weights can be achieved by those functions of Construction 2 satisfying that $g_1+g_2$ is with two monomials, which is true for $m=4$. However, for $m=5$, the weight 214 cannot be achieved by such functions. If further study shows that for $m\ge 6$ all the desired weights can be achieved by those functions such that $g_1+g_2$ is with two monomials, then by using Lemmas 3 and 5, one may prove the conjecture and determine the weight spectra of all codes $RM(m-c,m)$ for $m\ge 2c$, which would be a milestone in the research on Reed--Muller codes. However, if unlucky, some weights cannot be achieved, we then need to investigate the weights of those functions with four monomials and try to deduce an elegant formula. That is, a formula for $wt(f)$ with
\[
f=x_{1}x_{2}\cdots x_{m}+x_{i_1}\cdots x_{i_m}+x_{i_{m+1}}\cdots x_{i_{2m}}+x_{i_{2m+1}}\cdots x_{i_{3m}}\in B_{2m},
 \]
where $\{i_1,i_2,\cdots,i_{3m}\}\subseteq \{1,2,\cdots,2m\}$. If we succeed, then it could still be feasible to prove Conjecture 2.
\section{Conclusion}
Determining the weight distributions and the weight spectra of the Reed-Muller codes $RM(r,m)$ is a challenging task. The weight spectra of the Reed--Muller codes $RM(r,m)$ were unknown for $r=3,...,m-5$. In 2024, Carlet determined the weight spectrum of $RM(m-5,m)$ for $m\ge10$ using the Maiorana--McFarland construction, where he tried to extend the result to $RM(m-6,m)$, but not succeeded. In this paper, we propose a novel way of constructing Reed--Muller codewords and determine the weight spectrum of $RM(m-6,m)$ for $m\ge12$, which gives a positive answer to an open question on the weight spectrum of $RM(m-c,m)$ for $c=6$. 

We propose a construction based on the concatenation of four functions which can provide many weights, and put forward a conjecture on that construction. We verify the conjecture for some cases, and it seems to be feasible to prove it. Further study may verify the conjecture for $RM(7,14)$ and determine the weight spectrum of $RM(m-7,m)$. If the conjecture is true, then that open question can be completely solved, which would be a milestone in the research on Reed-Muller codes.

\section*{Acknowledgment}

The authors would like to thank the financial support from the National Natural Science Foundation of China (Grant 62172230) and Natural Science Foundation of Jiangsu
Province (No. BK20201369). We also thank the anonymous reviewers, whose comments helped improving the paper.





\end{document}